\documentclass[12pt]{article}
\usepackage{amssymb,amsmath,epsfig}
\begin{document}
\title{\bf Exact Solutions of Bianchi Types $I$ and $V$ Spacetimes in $f(R)$ Theory of Gravity}

\author{M. Sharif \thanks{msharif@math.pu.edu.pk} and M. Farasat
Shamir \thanks{frasat@hotmail.com}\\\\
Department of Mathematics, University of the Punjab,\\
Quaid-e-Azam Campus, Lahore-54590, Pakistan.}

\date{}

\maketitle
\begin{abstract}
In this paper, the crucial phenomenon of the expansion of the
universe has been discussed. For this purpose, we study the vacuum
solutions of Bianchi types $I$ and $V$ spacetimes in the framework
of $f(R)$ gravity. In particular, we find two exact solutions in
each case by using the variation law of Hubble parameter. These
solutions correspond to two models of the universe. The first
solution gives a singular model while the second solution provides a
non-singular model. The physical behavior of these models is
discussed. Moreover, the function of the Ricci scalar is evaluated
for both the models in each case.
\end{abstract}


\section{Introduction}

The accelerating expansion of the universe has attracted much
attention in recent years. Why the universe is expanding at an
increasing rate and spreading its contents over greater dimensions
of space? It is the most interesting question in the modern day
cosmology. There is another issue of dark energy and dark matter
which has been discussed widely. Einstein gave the concept of dark
energy by introducing the cosmological constant. Later, he remarked
that the introduction of the cosmological term was the biggest
blunder of his life.

Carmeli \cite{1} predicted for the first time in $1996$ that the
universe expansion was accelerating. Later, the supernova
experiments \cite{2} became known and it was found that major part
of the universe was filled with dark matter and dark energy. The
results from these experiments provided evidences for the
accelerating expansion of the universe. The cosmological constant
is considered as one of the candidate responsible for the dark
energy. Modified theories of gravity provide an alternative
approach to study the universe.

There are some useful aspects \cite{02} of modified theories of
gravity. Modified gravity gives an easy unification of early time
inflation and late time acceleration. It provides a natural
gravitational alternative to dark energy. The explanation of cosmic
acceleration is obtained just by introducing the term $1/R$ which is
essential at small curvatures. The modified gravity also describes
the transition phase of the universe from deceleration to
acceleration. It can be used for the explanation of hierarchy
problem in high energy physics. The $f(R)$ theory of gravity is one
of the modified theories which is considered most suitable due to
cosmologically important $f(R)$ models. These models consist of
higher order curvature invariants as functions of the Ricci scalar.
Viable $f(R)$ gravity models \cite{002} have been proposed which
show the unification of early-time inflation and late-time
acceleration. The problem of dark matter can also be addressed by
using viable $f(R)$ gravity models.

Singularity has been an important issue in General Relativity
(GR). The occurrence of spacetime singularity is a general feature
of any cosmological model under some reasonable conditions. It may
be possible to avoid these undesired singularities in the context
of modified theories. Kanti et al. \cite{3} showed that there did
not exist any cosmological singularity by considering higher order
curvature terms. Nojiri and Odintsov \cite{03} proposed some
realistic singularity free models in modified $f(R)$ gravity.
Bamba et al. \cite{003} proved that non-minimal gravitational
coupling can remove the finite-time future singularity in modified
gravity. Thus there is a strong reason to study solutions of the
field equations in modified theories to address the problems like
dark energy and singularity.

Weyl \cite{4} and Eddington \cite{5} studied $f(R)$ actions in
$1919$ and $1922$ respectively. Buchdahl \cite{6} explored these
actions in the context of non-singular oscillating cosmologies.
Cognola et al. \cite{07} investigated $f(R)$ gravity at one-loop
level in de-Sitter universe. It was found that one-loop effective
action can be useful for the study of constant curvature black hole
nucleation rate. Spherical symmetry is the most closest approach to
the nature because one can compare the results from the solar system
observations. Thus the most commonly explored exact solutions in
$f(R)$ gravity are the spherically symmetric solutions.
Multam$\ddot{a}$ki and Vilja \cite{7} studied spherically symmetric
vacuum solutions for the first time in this theory. The same authors
\cite{8} also investigated the perfect fluid solutions and showed
that pressure and density did not uniquely determine $f(R)$.
Capozziello et al. \cite{9} explored spherically symmetric solutions
of $f(R)$ theories of gravity via the Noether symmetry approach.
Hollenstein and Lobo \cite{10} analyzed the exact solutions of
static spherically symmetric spacetimes in $f(R)$ gravity coupled to
non-linear electrodynamics.

Cylindrical symmetry is next to spherical symmetry which may be used
to study the exact solutions of the field equations in $f(R)$
gravity. Azadi et al. \cite{11} studied cylindrically symmetric
vacuum solutions in metric $f(R)$ theory of gravity. Momeni
\cite{12} extended this work to the general cylindrically symmetric
solution. Recently, we have explored static plane symmetric vacuum
solutions \cite{13} in $f(R)$ theory of gravity. The field equations
are solved using the assumption of constant scalar curvature which
may be zero or non-zero. However, no attempt has been made so far
for solutions with non-constant scalar curvature.

The accelerating expansion of the universe can be studied using
Bianchi types $I$ and $V$ spacetimes which are the generalization of
FRW spacetimes. Due to the spatially homogeneous and isotropic
nature, many authors \cite{013}-\cite{018} have studied these
spacetimes in different contexts. Berman \cite{019} introduced a
different method to solve the field equations by using the variation
law of Hubble's parameter. The main feature of the variation law is
that it gives the constant value of deceleration parameter. Using
this law, Singh et al. \cite{020} explored perfect fluid solutions
of Bianchi type $V$ spacetime. Recently, Kumar and Singh \cite{021}
studied solutions of the field equations in the presence of perfect
fluid using Bianchi type $I$ spacetime in GR. The same authors
\cite{022} investigated perfect fluid solutions using Bianchi type
$I$ spacetime in scalar-tensor theory.

In this paper, we focuss our attention to explore the vacuum
solutions of Bianchi types $I$ and $V$ spacetimes in $f(R)$ theories
of gravity using metric approach. The paper is organized as follows:
In section \textbf{2}, we give a brief introduction about the field
equations in the context of $f(R)$ gravity. Sections \textbf{3} and
\textbf{4} are used to find exact vacuum solutions and the
singularity analysis of these solutions. In the last section, we
summarize and conclude the results.

\section{$f(R)$ Gravity Formalism}

The $f(R)$ theory of gravity is the generalization of GR. The
action for this theory is given by
\begin{equation}\label{1}
S=\int\sqrt{-g}(\frac{1}{16\pi{G}}f(R)+L_{m})d^4x.
\end{equation}
Here $f(R)$ is a general function of the Ricci scalar and $L_{m}$ is
the matter Lagrangian. It is noted that this action is obtained just
by replacing $R$ by $f(R)$ in the standard Einstein-Hilbert action.
The corresponding field equations are found by varying the action
with respect to the metric $g_{\mu\nu}$
\begin{equation}\label{2}
F(R)R_{\mu\nu}-\frac{1}{2}f(R)g_{\mu\nu}-\nabla_{\mu}
\nabla_{\nu}F(R)+g_{\mu\nu}\Box F(R)=\kappa T_{\mu\nu},
\end{equation}
where
\begin{equation}\label{3}
F(R)\equiv df(R)/dR,\quad\Box\equiv\nabla^{\mu}\nabla_{\mu},
\end{equation}
$\nabla_{\mu}$ is the covariant derivative and $T_{\mu\nu}$ is the
standard matter energy-momentum tensor derived from the Lagrangian
$L_m$. These are the fourth order partial differential equations in
the metric tensor. The fourth order is due to the last two terms on
the left hand side of the equation. If we take $f(R)=R$, these
equations reduce to the field equations of GR.

Now contracting the field equations, it follows that
\begin{equation}\label{4}
F(R)R-2f(R)+3\Box F(R)=\kappa T
\end{equation}
and in vacuum, we have
\begin{equation}\label{5}
F(R)R-2f(R)+3\Box F(R)=0.
\end{equation}
This gives an important relationship between $f(R)$ and $F(R)$ which
will be used to simplify the field equations and to evaluate $f(R)$.

\section{Exact Bianchi Type $I$ Solutions}

Here we shall find exact solutions of Bianchi I spacetime in
$f(R)$ gravity. For the sake of simplicity, we take the vacuum
field equations.

\subsection{Bianchi Type $I$ Spacetime}

The line element of the Bianchi type $I$ spacetime is given by
\begin{equation}\label{6}
ds^{2}=dt^2-A^2(t)dx^2-B^2(t)dy^2-C^2(t)dz^2,
\end{equation}
where $A,~B$ and $C$ are cosmic scale factors. The corresponding
Ricci scalar becomes
\begin{equation}\label{7}
R=-2[\frac{\ddot{A}}{A}+\frac{\ddot{B}}{B}+\frac{\ddot{C}}{C}
+\frac{\dot{A}\dot{B}}{AB}+\frac{\dot{B}\dot{C}}{BC}+\frac{\dot{C}\dot{A}}{CA}],
\end{equation}
where dot represents derivative with respect to $t$.

We define the average scale factor $a$ as
\begin{equation}\label{8}
a=\sqrt[3]{ABC}
\end{equation}
and the volume scale factor as
\begin{equation}\label{08}
V=a^3=ABC.
\end{equation}
We also define the generalized mean Hubble parameter $H$ in the form
\begin{equation}\label{008}
H=\frac{1}{3}(H_1+H_2+H_3),
\end{equation}
where
$H_1=\frac{\dot{A}}{A},~H_2=\frac{\dot{B}}{B},~H_3=\frac{\dot{C}}{C}$
are the directional Hubble parameters in the directions of $x,~y$
and $z$ axis respectively. Using Eqs.(\ref{8})-(\ref{008}), we
obtain
\begin{equation}\label{0008}
H=\frac{1}{3}\frac{\dot{V}}{V}=\frac{1}{3}(H_1+H_2+H_3)=\frac{\dot{a}}{a}.
\end{equation}

It follows from Eq.(\ref{5}) that
\begin{equation}\label{9}
f(R)=\frac{3\Box F(R)+F(R)R}{2}.
\end{equation}
Inserting this value of $f(R)$ in the vacuum field equations, we
have
\begin{equation}\label{10}
\frac{F(R)R_{\mu\nu}-\nabla_{\mu}\nabla_{\nu}F(R)}{g_{\mu\nu}}
=\frac{F(R)R-\Box F(R)}{4}.
\end{equation}
Since the metric (\ref{6}) depends only on $t$, one can view
Eq.(\ref{10}) as the set of differential equations for $F(t)$,
$A,~B$ and $C$. It follows from Eq.(\ref{10}) that the combination
\begin{equation}\label{11}
A_{\mu}\equiv\frac{F(R)R_{\mu\mu}-\nabla_{\mu}\nabla_{\mu}
F(R)}{g_{\mu\mu}}
\end{equation}
is independent of the index $\mu$ and hence $A_{\mu}-A_{\nu}=0$ for
all $\mu$ and $\nu$. Consequently, $A_0-A_1=0$ gives
\begin{equation}\label{12}
-\frac{\ddot{B}}{B}-\frac{\ddot{C}}{C}
+\frac{\dot{A}\dot{B}}{AB}+\frac{\dot{C}\dot{A}}{CA}
+\frac{\dot{A}\dot{F}}{AF}-\frac{\ddot{F}}{F}=0.
\end{equation}
Also, $A_{0}-A_{2}=0$ and $A_{0}-A_{3}=0$ yield respectively
\begin{eqnarray} \label{13}
-\frac{\ddot{A}}{A}-\frac{\ddot{C}}{C}
+\frac{\dot{A}\dot{B}}{AB}+\frac{\dot{B}\dot{C}}{BC}
+\frac{\dot{B}\dot{F}}{BF}-\frac{\ddot{F}}{F}=0,\\\label{14}
-\frac{\ddot{A}}{A}-\frac{\ddot{B}}{B}
+\frac{\dot{B}\dot{C}}{BC}+\frac{\dot{C}\dot{A}}{CA}
+\frac{\dot{C}\dot{F}}{CF}-\frac{\ddot{F}}{F}=0.
\end{eqnarray}
Thus we get three non-linear differential equations with four
unknowns namely $A,~B,~C$ and $F$. The solution of these equations
can be found by following the approach of Saha \cite{14}.

\subsection{Solution of the Field Equations}

Subtracting Eq.(\ref{13}) from Eq.(\ref{12}), Eq.(\ref{14}) from
Eq.(\ref{13}) and Eq.(\ref{14}) from Eq.(\ref{11}), we get
respectively
\begin{eqnarray}\label{015}
\frac{\ddot{A}}{A}-\frac{\ddot{B}}{B}
+\frac{\dot{C}}{C}(\frac{\dot{A}}{A}-\frac{\dot{B}}{B})
+\frac{\dot{F}}{F}(\frac{\dot{A}}{A}-\frac{\dot{B}}{B})=0,\\\label{016}
\frac{\ddot{B}}{B}-\frac{\ddot{C}}{C}
+\frac{\dot{A}}{A}(\frac{\dot{B}}{B}-\frac{\dot{C}}{C})
+\frac{\dot{F}}{F}(\frac{\dot{B}}{B}-\frac{\dot{C}}{C})=0,\\\label{017}
\frac{\ddot{A}}{A}-\frac{\ddot{C}}{C}
+\frac{\dot{B}}{B}(\frac{\dot{A}}{A}-\frac{\dot{C}}{C})
+\frac{\dot{F}}{F}(\frac{\dot{A}}{A}-\frac{\dot{C}}{C})=0.
\end{eqnarray}
These equations imply that
\begin{eqnarray}\label{15}
\frac{B}{A}=d_1\exp[{c_1\int\frac{dt}{a^3F}}],\\\label{16}
\frac{C}{B}=d_2\exp[{c_2\int\frac{dt}{a^3F}}],\\\label{17}
\frac{A}{C}=d_3\exp[{c_3\int\frac{dt}{a^3F}}],
\end{eqnarray}
where $c_1,~c_2,~c_3$ and $d_1,~d_2,~d_3$ are constants of
integration which satisfy the relation
\begin{equation}\label{18}
c_1+c_2+c_3=0,\quad d_1d_2d_3=1.
\end{equation}
Using Eqs.(\ref{15})-(\ref{17}), we can write the metric functions
explicitly as
\begin{eqnarray}\label{19}
A=ap_1\exp[{q_1\int\frac{dt}{a^3F}}],\\\label{20}
B=ap_2\exp[{q_2\int\frac{dt}{a^3F}}],\\\label{21}
C=ap_3\exp[{q_3\int\frac{dt}{a^3F}}],
\end{eqnarray}
where
\begin{equation}\label{22}
p_1=({d_1}^{-2}{d_2}^{-1})^{\frac{1}{3}},\quad
p_2=(d_1{d_2}^{-1})^{\frac{1}{3}},\quad
p_3=(d_1{d_2}^2)^{\frac{1}{3}}
\end{equation}
and
\begin{equation}\label{23}
q_1=-\frac{2c_1+c_2}{3},\quad q_2=\frac{c_1-c_2}{3},\quad
q_3=\frac{c_1+2c_2}{3}.
\end{equation}
Notice that $p_1,~p_2,~p_3$ and $q_1,~q_2,~q_3$ also satisfy the
relation
\begin{equation}\label{24}
p_1p_2p_3=1,\quad q_1+q_2+q_3=0.
\end{equation}

Now we use the power law assumption to solve the integral part in
the above equations. The power law relation between scale factor and
scalar field has already been used by Johri and Desikan \cite{0015}
in the context of Robertson-Walker Brans-Dicke models. However, in a
recent paper \cite{0016}, Kotub Uddin et al. have established a
result in the context of $f(R)$ gravity which shows that
\begin{equation*}
F\propto a^m,
\end{equation*}
where $m$ is an arbitrary constant. Thus using power law relation
between $F$ and $a$, we have
\begin{equation}\label{25}
F=ka^m,
\end{equation}
where $k$ is the constant of proportionality and $m$ is any
integer. The deceleration parameter $q$ in cosmology is the
measure of the cosmic acceleration of the universe expansion and
is defined as
\begin{equation}\label{26}
q=-\frac{\ddot{a}a}{\dot{a}^2}.
\end{equation}
It is mentioned here that the negative sign and the name
"deceleration parameter" are historical. Initially, $q$ was supposed
to be positive but recent observations from the supernova
experiments suggest that it is negative. Thus the behavior of the
universe models depend upon the sign of $q$. The positive
deceleration parameter corresponds to a decelerating model while the
negative value provides inflation. We also use a well-known relation
\cite{15} between the average Hubble parameter $H$ and average scale
factor $a$ given as
\begin{equation}\label{27}
H=la^{-n},
\end{equation}
where $l>0$ and $n\geq0$. This is an important relation because it
gives the constant value of the deceleration parameter.

Using Eqs.(\ref{0008}) and (\ref{27}), it follows that
\begin{equation}\label{28}
\dot{a}=la^{1-n}
\end{equation}
and consequently the deceleration parameter turns out to be
\begin{equation}\label{29}
q=n-1
\end{equation}
which is obviously a constant. Integrating Eq.(\ref{28}), it follows
that
\begin{equation}\label{30}
a=(nlt+k_1)^{\frac{1}{n}},\quad n\neq0
\end{equation}
and
\begin{equation}\label{31}
a=k_2\exp(lt),\quad n=0,
\end{equation}
where $k_1$ and $k_2$ are constants of integration. Thus we obtain
two values of the average scale factor that correspond to two
different models of the universe.

\subsection{Singularity Analysis}

The Riemann tensor is useful to determine whether a singularity is
essential or coordinate. If the curvature becomes infinite at a
certain point, then the singularity is essential. By constructing
scalars from the Riemann tensor, it can be checked whether they
become infinite somewhere or not. It is obvious that infinitely many
scalars can be constructed from the Riemann tensor. However,
symmetry considerations can be used to show that there are only a
finite number of independent scalars. All others can be expressed in
terms of these. In a four-dimensional Riemann spacetime, there are
only $14$ independent curvature invariants. Some of these are
\begin{eqnarray*}
R_1=R=g^{ab}R_{ab},\quad R_2=R_{ab}R^{ab},\quad
R_3=R_{abcd}R^{abcd},\quad R_4=R^{ab}_{cd}R^{cd}_{ab}.
\end{eqnarray*}
Here we give analysis for the first invariant commonly known as the
Ricci scalar. For Bianchi type $I$ spacetime, it is given by
Eq.(\ref{7}). For a special case when $m=-2$, it follows from
Eq.(\ref{25}) that
\begin{equation}\label{33}
F=ka^{-2}.
\end{equation}
After some manipulations, we can write
\begin{equation}\label{32}
R_1=-2[\frac{3k^2(a\ddot{a}+\dot{a}^2)-(q_1q_2+q_2q_3+q_3q_1)}{k^2a^2}]
\end{equation}
which shows that singularity occurs at $a=0$.

\subsection{Model of the Universe when $n\neq0$.}

Now we discuss model of the universe when $n\neq0$, i.e.,
$a=(nlt+k_1)^{\frac{1}{n}}$. For this model, $F$ becomes
\begin{equation}\label{34}
F=k(nlt+k_1)^{-\frac{2}{n}}.
\end{equation}
Using this value of $F$ in Eqs.(\ref{19})-(\ref{21}), the metric
coefficients $A,~B$ and $C$ turn out to be
\begin{eqnarray}\label{35}
A&=&p_1(nlt+k_1)^{\frac{1}{n}}\exp[\frac{q_1(nlt+k_1)^
{\frac{n-1}{n}}}{kl(n-1)}],\quad n\neq1\\\label{36}
B&=&p_2(nlt+k_1)^{\frac{1}{n}}\exp[\frac{q_2(nlt+k_1)^
{\frac{n-1}{n}}}{kl(n-1)}],\quad n\neq1\\\label{37}
C&=&p_3(nlt+k_1)^{\frac{1}{n}}\exp[\frac{q_3(nlt+k_1)^
{\frac{n-1}{n}}}{kl(n-1)}],\quad n\neq1.
\end{eqnarray}
The directional Hubble parameters $H_i$ ($i=1,2,3$) take the form
\begin{equation}\label{38}
H_i=\frac{l}{nlt+k_1}+\frac{q_i}{k(nlt+k_1)^{\frac{1}{n}}}.
\end{equation}
The mean generalized Hubble parameter becomes
\begin{equation}\label{39}
H=\frac{l}{nlt+k_1}
\end{equation}
while the volume scale factor turns out to be
\begin{equation}\label{40}
V=(nlt+k_1)^\frac{3}{n}.
\end{equation}
Moreover, the function of Ricci scalar, $f(R)$, can be found by
using Eq.(\ref{9})
\begin{equation}\label{41}
f(R)=\frac{k}{2}(nlt+k_1)^\frac{-2}{n}R+3kl^2(n-1)(nlt+k_1)^\frac{-2n-2}{n}.
\end{equation}
It follows from Eq.(\ref{32}) that
\begin{equation}\label{032}
R\equiv R_1=-2[3l^2(2-n)(nlt+k_1)^{-2}-
\frac{(q_1q_2+q_2q_3+q_3q_1)} {k^2}(nlt+k_1)^{\frac{-2}{n}}],
\end{equation}
which clearly indicates that $f(R)$ cannot be explicitly written in
terms of $R$. However, by inserting this value of $R$, $f(R)$ can be
written as a function of $t$, which is true as $R$ depends upon $t$.
For a special case when $n=\frac{1}{2}$, $f(R)$ turns out to be
\begin{eqnarray}\nonumber
f(R)&=&\frac{k}{2}[\frac{-9l^2\pm
\sqrt{81l^4+\frac{8(q_1q_2+q_2q_3+q_3q_1)}{k^2}R}}{2R}]^2R
\\&-& \frac{3kl^2}{2}[\frac{-9l^2\pm
\sqrt{81l^4+\frac{8(q_1q_2+q_2q_3+q_3q_1)}{k^2}R}}{2R}]^3.
\end{eqnarray}
This gives $f(R)$ only as a function of $R$.

\subsection{Model of the Universe when $n=0$.}

The average scale factor for this model of the universe is
$a=k_2\exp(lt)$ and hence $F$ takes the form
\begin{equation}\label{42}
F=\frac{k}{{k_2}^2}\exp(-2lt).
\end{equation}
Inserting this value of $F$ in Eqs.(\ref{19})-(\ref{21}), the metric
coefficients $A,~B$ and $C$ become
\begin{eqnarray}\label{43}
A&=&p_1k_2\exp(lt)\exp[-\frac{q_1\exp(-lt)}{klk_2}],\\\label{44}
B&=&p_2k_2\exp(lt)\exp[-\frac{q_2\exp(-lt)}{klk_2}],\\\label{37}
C&=&p_3k_2\exp(lt)\exp[-\frac{q_3\exp(-lt)}{klk_2}].
\end{eqnarray}
The directional Hubble parameters $H_i$ and the mean generalized
Hubble parameter will become
\begin{equation}\label{44}
H_i=l+\frac{q_i}{kk_2}\exp(-lt),\quad H=l.
\end{equation}
The volume scale factor turns out to be
\begin{equation}\label{46}
V={k_2}^3\exp(3lt)
\end{equation}
while $f(R)$ takes the form
\begin{equation}\label{47}
f(R)=\frac{k}{{2k_2}^2}\exp(-2lt)(R-6l^2).
\end{equation}
For this model, $R$ becomes
\begin{equation}\label{0032}
R=-2[6l^2-\frac{(q_1q_2+q_2q_3+q_3q_1)}{k^2{k_2}^2\exp(2lt)}].
\end{equation}
Here we can get the general function $f(R)$ in terms of $R$
\begin{equation}\label{047}
f(R)=\frac{k^3}{2(q_1q_2+q_2q_3+q_3q_1)}[R^2+6l^2R-72l^4]
\end{equation}
which corresponds to the general function $f(R)$ \cite{16},
\begin{equation}\label{0047}
f(R)=\sum a_n R^n,
\end{equation}
where $n$ may take the values from negative or positive.

\section{Exact Bianchi Type $V$ Solutions}

Here we shall find exact solutions of the Bianchi type $V$ spacetime
in $f(R)$ gravity for the vacuum field equations.

\subsection{Bianchi Type $V$ Spacetime}

The metric for the Bianchi type $V$ spacetime is
\begin{equation}\label{56}
ds^{2}=dt^2-A^2(t)dx^2-e^{2mx}[B^2(t)dy^2+C^2(t)dz^2],
\end{equation}
where $A,~B$ and $C$ are cosmic scale factors and $m$ is an
arbitrary constant. The corresponding Ricci scalar is
\begin{equation}\label{57}
R=-2[\frac{\ddot{A}}{A}+\frac{\ddot{B}}{B}+\frac{\ddot{C}}{C}-\frac{3m^2}{A^2}
+\frac{\dot{A}\dot{B}}{AB}+\frac{\dot{B}\dot{C}}{BC}+\frac{\dot{C}\dot{A}}{CA}].
\end{equation}
With the help of Eq.(\ref{11}), we can write
$A_0-A_1=0,~A_{0}-A_{2}=0$ and $A_{0}-A_{3}=0$ respectively as
\begin{eqnarray}\label{512}
-\frac{\ddot{B}}{B}-\frac{\ddot{C}}{C}-\frac{2m^2}{A^2}
+\frac{\dot{A}\dot{B}}{AB}+\frac{\dot{C}\dot{A}}{CA}
+\frac{\dot{A}\dot{F}}{AF}-\frac{\ddot{F}}{F}=0,\\ \label{513}
-\frac{\ddot{A}}{A}-\frac{\ddot{C}}{C}-\frac{2m^2}{A^2}
+\frac{\dot{A}\dot{B}}{AB}+\frac{\dot{B}\dot{C}}{BC}
+\frac{\dot{B}\dot{F}}{BF}-\frac{\ddot{F}}{F}=0,\\\label{514}
-\frac{\ddot{A}}{A}-\frac{\ddot{B}}{B}-\frac{2m^2}{A^2}
+\frac{\dot{B}\dot{C}}{BC}+\frac{\dot{C}\dot{A}}{CA}
+\frac{\dot{C}\dot{F}}{CF}-\frac{\ddot{F}}{F}=0.
\end{eqnarray}
The $01$-component can be written by using Eq.(\ref{2}) in the
following form
\begin{equation}\label{0513}
2\frac{\dot{A}}{A}-\frac{\dot{B}}{B}-\frac{\dot{C}}{C}=0.
\end{equation}
We solve these equations using the same procedure as for the Bianchi
type $I$ solutions.

\subsection{Solution of the Field Equations}

Here we get same equations Eqs.(\ref{015})-(\ref{017}) as obtained
previously with the difference of the constraint equations (using
Eq.(\ref{0513}))
\begin{equation}\label{0524}
p_1=1,\quad p_2={p_3}^{-1}=P,\quad q_1=0,\quad q_2=-q_3=Q.
\end{equation}
Consequently, the metric functions become
\begin{eqnarray}\label{0519}
A=a,\quad B=aP\exp[{Q\int\frac{dt}{a^3F}}],\quad
C=aP^{-1}\exp[{-Q\int\frac{dt}{a^3F}}].
\end{eqnarray}
Moreover, $R_1$ turns out to be
\begin{equation}\label{532}
R_1=-2[\frac{3k^2(a\ddot{a}+\dot{a}^2)+Q^2-3m^2k^2}{k^2a^2}]
\end{equation}
which also yields a singularity at $a=0$.

\subsection{Model of the Universe when $n\neq0$.}

Using the value of $F$ from Eq.(\ref{34}) in Eq.(\ref{0519}), the
metric coefficients $A,~B$ and $C$ become
\begin{eqnarray}\label{535}
A&=&(nlt+k_1)^{\frac{1}{n}},\\\label{536}
B&=&P(nlt+k_1)^{\frac{1}{n}}\exp[\frac{Q(nlt+k_1)^
{\frac{n-1}{n}}}{kl(n-1)}],\quad n\neq1\\\label{537}
C&=&P^{-1}(nlt+k_1)^{\frac{1}{n}}\exp[\frac{-Q(nlt+k_1)^
{\frac{n-1}{n}}}{kl(n-1)}],\quad n\neq1.
\end{eqnarray}
The directional Hubble parameters $H_1,~H_2$ and $H_3$ take the
following form
\begin{eqnarray}\label{538}
H_1&=&\frac{l}{nlt+k_1},\\\label{0538}
H_2&=&\frac{l}{nlt+k_1}+\frac{Q}{k(nlt+k_1)^{\frac{1}{n}}},\\\label{00538}
H_2&=&\frac{l}{nlt+k_1}-\frac{Q}{k(nlt+k_1)^{\frac{1}{n}}}.
\end{eqnarray}
Notice that the mean generalized Hubble parameter $H$, the volume
scale factor $V$ and $f(R)$ turn out to be the same as for the
Bianchi type $I$ spacetime.

\subsection{Model of the Universe when $n=0$.}

For this model, the metric coefficients $A,~B$ and $C$ turn out to
be
\begin{eqnarray}\label{543}
A&=&k_2\exp(lt),\\\label{544}
B&=&Pk_2\exp(lt)\exp[-\frac{Q\exp(-lt)}{klk_2}],\\\label{545}
C&=&P^{-1}k_2\exp(lt)\exp[\frac{Q\exp(-lt)}{klk_2}].
\end{eqnarray}
The directional Hubble parameters $H_1,~H_2$ and $H_3$ take the
form
\begin{eqnarray}\label{546}
H_1=l,\quad H_2=l+\frac{Q\exp(-lt)}{kk_2},\quad
H_3=l-\frac{Q\exp(-lt)}{kk_2}.
\end{eqnarray}
Here we also have the same mean generalized Hubble parameter $H$,
volume scale factor $V$ and $f(R)$ as given in
Eqs.(\ref{44})-(\ref{47}).

\section{Summary and Conclusion}

The main purpose of this paper is to discuss the well-known
phenomenon of the universe expansion in the context of $f(R)$
gravity. For this purpose, we have investigated exact solutions of
the Bianchi types $I$ and $V$ spacetimes using the vacuum field
equations. We have found two exact solutions for both spacetimes by
using the variation law of Hubble parameter. This yields constant
value of deceleration parameter. These solutions correspond to two
models of the universe. The first solution gives a singular model
with power law expansion and positive deceleration parameter while
the second solution provides a non-singular model with exponential
expansion and negative deceleration parameter. It is mentioned here
that the solutions for both spacetimes correspond to perfect fluid
solutions \cite{020, 021} in GR. We have also evaluated function of
the Ricci scalar, $f(R)$, for both models in each case. In
particular, the general function $f(R)$ includes squared power of
the Ricci scalar for the non-singular model. The physical behavior
of these models is given below.

First we discuss singular model of the universe. This model
corresponds to $n\neq0$ with average scale factor
$a=(nlt+k_1)^{\frac{1}{n}}$. It has a point singularity at $t\equiv
t_s=-\frac{k_1}{nl}$. The physical parameters $H_1,~H_2,~H_3$ and
$H$ are all infinite at this point but the volume scale factor
vanishes here. The function of the Ricci scalar, $f(R)$, is also
infinite while the metric functions $A,~B$ and $C$ vanish at this
point of singularity. Thus we can conclude from these observations
that the model starts its expansion with zero volume at $t=t_s$ and
it continues to expand for $0<n<1$.

The non-singular model of the universe corresponds to $n=0$ with
average scale factor $a=k_2\exp(lt)$. It is non-singular because
exponential function is never zero and hence there does not exist
any physical singularity for this model. The physical parameters
$H_1,~H_2,~H_3$ are all finite for all finite values of $t$. The
mean generalized Hubble parameter $H$ is constant while $f(R)$ is
also finite here. The metric functions $A,~B$ and $C$ do not vanish
for this model. The volume scale factor increases exponentially with
time which indicates that the universe starts its expansion with
zero volume from infinite past.

It would be worthwhile to study other Bianchi type spacetimes
especially by removing the vacuum condition. It would also be
interesting to explore Bianchi types $I$ and $V$ solutions with
perfect fluid. These are under progress.

\end{document}